\documentclass{article}

\usepackage{arxiv}

\usepackage[utf8]{inputenc} 
\usepackage[T1]{fontenc}    
\usepackage{hyperref}       
\usepackage{url}            
\usepackage{booktabs}       
\usepackage{amsfonts}       
\usepackage{nicefrac}       
\usepackage{microtype}      
\usepackage{lipsum}
\usepackage{graphicx}
\graphicspath{ {./images/} }

\title{Cooperation as a Black Box: Conceptual Fluctuation and Diagnostic Tools for Misalignment in MAS}

\author{Shayak Nandi \\  
  Grinnell College, \\
  Grinnell, IA 50112\\
  \texttt{nandisha@grinnell.edu} \\
   \And
Fernanda M. Eliott \\
  Grinnell College, \\
  Grinnell, IA 50112\\
  \texttt{eliottfe@grinnell.edu} \\
  }

\begin{document}
\maketitle
\begin{abstract}
Misalignment in Multi-Agent Systems (MAS) is frequently treated as a technical failure. Yet, issues may arise from the conceptual design phase, where semantic ambiguity and normative projection occur. The Rabbit–Duck illusion illustrates how perspective-dependent readings of agent behavior, such as the conflation of cooperation/coordination, can create epistemic instability; \textit{e.g.}, coordinated agents in cooperative Multi-Agent Reinforcement Learning (MARL) benchmarks being interpreted as morally aligned, despite being optimized for shared utility maximization only. Motivated by three drivers of meaning-level misalignment in MAS (coordination–cooperation ambiguity, conceptual fluctuation, and semantic instability), we introduce the \textit{Misalignment Mosaic}: a framework for diagnosing how misalignment emerges through language, framing, and design assumptions. The Mosaic comprises four components: 1. Terminological Inconsistency, 2. Interpretive Ambiguity, 3. Concept-to-Code Decay, and 4. Morality as Cooperation. Building on insights from the Morality-as-Cooperation Theory, we call for consistent meaning-level grounding in MAS to ensure systems function as intended: technically and ethically. This need is particularly urgent as MAS principles influence broader Artificial Intelligence (AI) workflows, amplifying risks in trust, interpretability, and governance. While this work focuses on the coordination/cooperation ambiguity, the Mosaic generalizes to other overloaded terms, such as alignment, autonomy, and trust.
\end{abstract}


\section{Interpretive Ambiguity as Alignment Risk}\label{sec_Intro}

Consider the following two situations: you notice a stranger struggling to carry a large box and offer help; you then share the physical effort to transport it to a goal location. Later at home, you and a partner assemble a jigsaw puzzle, coordinating your actions toward a shared goal. What distinguishes cooperation from coordination in these everyday interactions? And how consistently are such distinctions, or ambiguity, addressed in MAS research?

Challenges in consistently separating these \textit{terms}\footnote{We anchor ``concept'' to a word's associated abstract ideas \cite{concept_def}, whereas ``term'' to its more precise or technical meaning \cite{term_def}.} reflects a systemic conflation, with real structural consequences. Coordinated behavior is often misread as cooperation, leading to assumptions of moral alignment or shared intent where none exist. This interpretive jump, rooted in the observer rather than the agent, distorts how MAS are evaluated, trusted, and governed, as cooperation has ties to morality \cite{tomasello2013origins}: when coordination is interpreted as cooperation, systems may be evaluated on what we assume they mean instead of on what they do.

Coordinating agents under uncertainty is a core topic in MAS, rooted in distributed systems~\cite{lynch1996distributed}, game theory~\cite{chakravarty2014course}, and machine learning~\cite{lanctot2017unified}. Similarly, cooperation in Multi-Agent Learning intersects disciplines such as economics, social sciences, and evolutionary biology~\cite{du2023review}. Within MAS, diverse research traditions (\textit{e.g.}, Reinforcement Learning (RL), normative agent design) bring various assumptions about what \textit{cooperation} entails and how it should be modeled. While terminological flexibility enables methodological innovation, the coordination/cooperation ambiguity is a tricky case because it introduces moral interpretations into the design and evaluation of MAS.

As with our opening example, the distinction between coordination/cooperation is often context-dependent, shaped by factors such as the task, environment, and agents' affordances~\cite{gibson1966senses} \cite{gibson2014ecological}. Contextual variability has likely contributed to the interchangeable use of these terms across studies; however, this ambiguity introduces unintended consequences in MAS research and applications, particularly as other domains build upon MAS foundations. In fact, similar risks arise in emerging Generative AI systems that integrate agent-like structures and decision processes. As these systems scale, ambiguity becomes operational, with human observers projecting intent, values, or goals onto structural outputs. Thus, this paper invites MAS researchers to audit how semantic ambiguity and conceptual assumptions migrate into system design and deployment (often with normative implications beyond the technical layer).

\begin{figure*}[h]
  \centering
  \includegraphics[width=0.65\linewidth]{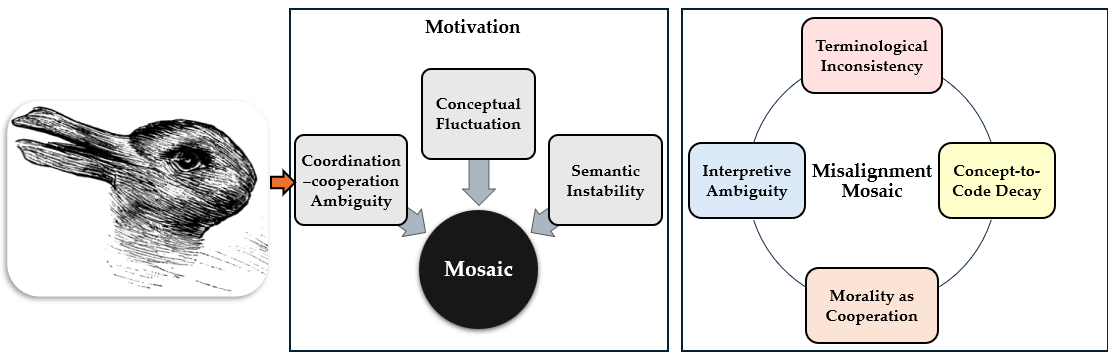}
  \caption{Left: The Rabbit–Duck illusion (made famous by Wittgenstein~\cite{wittgenstein}), illustrating how observer perspective influences MAS (e.g., is behavior seen as coordination or cooperation?). Misalignment Mosaic's three motivations at the Center, and at Right, a visual depiction (a diagnostic framework for tracing semantic and interpretive misalignment).}
  \label{fig:duck} \label{fig:map}  
\end{figure*}

The coordination/cooperation ambiguity mirrors the Rabbit– Duck illusion (see Figure~\ref{fig:duck}), where perception shifts without changes to the underlying image. Similarly, MAS behavior may be framed as coordination or cooperation depending on the observer's perspective and assumptions. Then, what initially appears as ``coordination'' may, upon reinterpretation, reveal underlying cooperative dynamics, just as shifting one's perception of the image reveals the duck within the rabbit. This ambiguity deepens when viewed through the lens of Braitenberg's thought experiments~\cite{braitenberg1986vehicles}: the \textit{vehicles} operate under simple rules, yet observers easily project complex intentions onto them (see~\cite{prescott2019synthetic}, \cite{shaikh2020braitenberg}, and ~\cite{Vehicles2023} for more perspectives on synthetic psychology and the vehicles). Such perspective-dependent interpretations complicate the design, analysis, and evaluation of MAS, as system behavior becomes linked with the observer's framing, rather than reflecting only the system's architecture.

 Therefore, we claim that a more intentional effort to distinguish coordination from cooperation is overdue, along with shared terminology, benchmarks~\cite{hanks1993benchmarks}, and interpretive literacy to scaffold rigorous MAS design and evaluation. To show how the problem extends beyond technical misalignment, we introduce the Misalignment Mosaic: a framework for diagnosing how semantic and interpretive ambiguity migrate from language to system design and ultimately to evaluation failures. 
 
 Rather than compressing methodological diversity, the Mosaic fosters shared vocabulary and practices. It complements recent alignment research by focusing on meaning itself, not only on how agents behave, but also on how their behavior is framed and morally interpreted in MAS. This semantic gap remains structurally unaddressed in MAS, despite its growing relevance in public-facing systems.

The term ``mosaic'' illustrates that MAS misalignments occur across shapes, scales, and contexts, collectively forming a complex interpretive heterogeneity — a phenomenon of \textit{alignment magnification}~\cite{dung2023current}. In its current structure, the Misalignment Mosaic comprises four diagnostic components, detailed in Section~\ref{sec_mosaic} and shown in Figure~\ref{fig:map}. Our main contributions are: 

\textbf{(1)} We identify three interconnected drivers of interpretive misalignment in MAS: a) \textbf{Coordination/cooperation ambiguity}, where these terms are used interchangeably, blurring conceptual boundaries; b) \textbf{Conceptual fluctuation}, where key terms like \textit{cooperation} acquire shifting meanings across system architecture, agent behavior, evaluation, and observer framing; and c) \textbf{Semantic instability}, where the anchoring of key terms becomes fragile or inconsistent, allowing meaning to drift across subfields, implementations, or observer perspectives. These dynamics systematically enable normative overreadings, in which coordination can be misread as cooperation and cooperation as moral alignment, producing interpretive, design-level, evaluative, and ultimately ethical risks.

\textbf{(2)}  We introduce the \textit{Misalignment Mosaic}: a framework for diagnosing meaning-level misalignment in MAS. It comprises four components: 1. Terminological Inconsistency, 2. Interpretive Ambiguity, 3. Concept-to-Code Decay, and 4. Morality as Cooperation. Together, these components help trace how misalignment arises not only from agent architecture or policy but from linguistic, epistemic, and framing assumptions embedded in MAS.

Variations in how \textit{coordination} and \textit{cooperation} are defined influence theoretical models, experimental design, and interdisciplinary communication. Inconsistencies are particularly problematic in multidisciplinary and multilingual research settings, where different conceptual models silently shape interpretation and evaluation. Rather than prescribing fixed definitions, we argue that diagnosing semantic instability and interpretive ambiguity as structural risks is appropriate, as these risks intensify as MAS scale into institutional and societal domains, where ambiguity itself becomes operational and must be recognized before alignment can be assessed or governed.
 
This work is organized as follows: Section~\ref{sec_Intro} introduces and motivates our ideas. Section~\ref{sec_background} presents the coordination/cooperation ambiguity, followed by Section~\ref{sec_mosaic}, which details the Misalignment Mosaic. In Section~\ref{sec_discussion}, we discuss our ideas, followed by a conclusion in Section~\ref{sec_conclusion}.

\section{Cooperation and Meaning-level  Misalignment}\label{sec_background}

\subsection{Semantic Instability and Everyday Settings}

Familiar settings exemplify the similarity between \textit{coordination} and \textit{cooperation}. Both structure how humans divide labor, respond to requests, and signal intent: ``Thanks for your cooperation,'' or ``Let's coordinate on this.'' Yet when imported into MAS design or evaluation frameworks, their meanings become unstable: is \textit{cooperation} defined by agent behavior, internal intent, task structure, or outcome? What distinguishes cooperation from strategic alignment, sacrifice, or simple compliance? Standard definitions provide little clarity: \textit{coordination} is often framed as ``the process of organizing people or groups so that they work together properly and well'' or ``the harmonious functioning of parts for effective results''~\cite{coordination_def}, while \textit{cooperation} implies shared effort or helpfulness~\cite{coop_def}. 

\textbf{Why does it matter?} Semantic instability can propagate misunderstandings and normative projection as systems scale in complexity, especially when more advanced frameworks are built on top of them. If key terms are variously defined at lower levels, the resulting inconsistencies will spread to more sophisticated applications. In fact, Barad~\cite{barad2007meeting} argues that matter and meaning are inextricably linked, emphasizing that scientific inquiry must account for the inseparability of epistemological, ontological, and ethical considerations.

Indeed, Langer {\it et al.}~\cite{langer2022look} investigate if terminological differences affect people's 1) Perceptions of properties of algorithmic decision-making systems, 2) Evaluations of systems in application contexts, and 3) The robustness and replicability of human-computer-interaction research. Their results emphasize the importance of carefully selecting terminology, as it can significantly bias users' interpretations and judgments of algorithmic systems -- the Citizen Science\cite{eitzel2017citizen} examines similar ideas. Similarly, aiming to build systems that can describe what they perceive using the users' terms, Erculiani and colleagues \cite{erculiani2023egocentric} claim that object recognition systems should follow a process that resembles the human when thinking of an object associated with a certain concept.

In MAS, agents may coordinate without explicit cooperation or cooperate without centralized coordination. \textit{Cooperation} itself is overloaded: spanning strategic alliances, sacrificial gestures, or even tacit collusion. Related terms, such as collaboration, assistance, opposite coordination, required cooperation \cite{zhang2016formal}, collusion, coalition, cooperative-competitive dynamics \cite{tuyls2012multiagent}, and advantage alignment \cite{duque2024advantage}, further complicate the landscape. Each variation implies different agent behaviors, moral interpretations, and normative assessments, underscoring the importance of clarity in MAS design and evaluation. Similar issues extend beyond MAS; in fact, Mesly et al. ~\cite{mesly2022terminology} analyze how economic academics inconsistently define ``predation'' and sometimes even erroneously or too narrowly.

To surface other issues at play, we raise the following questions for reflection:
\begin{enumerate}    
    \item Is coordination a structural prerequisite for cooperation\cite{huhns1999multiagent}? (Do different definitions change the answer?)      
    \item Do \textit{explicit instruction} and/or \textit{internally inferred} goal to \textit{cooperate} with others presuppose \textit{coordination}?      
    \item Are coordination and cooperation externally legible, or should cooperation be somewhat internal, \textit{i.e.}, linked to intention or value models? 
    \item Can cooperation be evaluated without invoking normative readings and moral projection (more in Section~\ref{sec_mosaic})? 
    \item Is \textit{collaboration} a subset of \textit{cooperation}? If so, what does that imply about hierarchical structures across different definitions of cooperation?
\end{enumerate}

We do not answer these questions; on the contrary, we hope it helps illustrate the relevance of meaning-level misalignment and how it can spread to technical consequences.

\subsection{Coordination vs. Cooperation in the MAS Literature}

Malone and Crowston~\cite{malone1990coordination} define \textit{coordination} as ``the act of managing interdependencies between activities performed to achieve a goal''. Building on that, agent coordination can be understood as the ``ability to manage the interdependencies of activities between agents''~\cite{consoli2007link}. In contrast, agent cooperation involves the ``process used for an agent to voluntarily enter a relationship with another to achieve a system derived goal''~\cite{consoli2007link}. Coordination often highlights agents' inter-dependencies management or task-level alignment, whereas cooperation often invokes common goals, altruistic behavior, and even moral norms (the ``Morality-as-Cooperation'' Theory). This distinction is further illustrated by Jensen {\emph et al.}~\cite{jensen2015testing}, who examine how people use accents of unfamiliar individuals to make social decisions through economic games that measure interpersonal trust, generosity, and coordination. Their results suggest that humans use ethnic markers of unfamiliar individuals to coordinate behavior rather than to cooperate.

Upon closer examination of the MAS literature, it becomes clear that the definition of \textit{cooperation} varies significantly. For example, Tan~\cite{tan1993multi} contrasts independent learners with cooperative agents, defining \emph{cooperation} as agents sharing episodes, learned policies, or instantaneous information, whereas others may take a different approach. As in \cite{matignon2012independent} (following \cite{boutilier1999sequential}), ``cooperative robots'' and ``fully cooperative MAS learning algorithms'' are settings where agents share common interests or the same utility function. In such cases, there is a correspondence between each agent's achievement and the group's, and therefore, the learning goal is defined as maximizing the common discounted return. Or, as \cite{boutilier1999sequential} puts it, fully cooperative MAS in which we assume that it is possible to set a common coordination mechanism and that agents do not have a reason to deliberate strategically. These variations highlight the terminological inconsistencies in MAS research, where cooperation can mean anything from simple information sharing to fully shared objectives and policies. Addressing these distinctions is essential for developing a coherent theoretical foundation for MAS.

 While the empirical studies above tackle different challenges in MAS, we argue that three interconnected drivers (coordination/cooperation ambiguity, conceptual fluctuation, and semantic instability) must be addressed before empirical methodologies can be effectively compared. Those issues lead to terminological inconsistencies in MAS research, resulting in conceptual fragmentation and making it difficult to establish common frameworks and methodologies. We identify a few factors that compound the three interconnected drivers by making them harder to detect, evaluate, or resolve:
\begin{enumerate}
    \item \textbf{Lack of uniformity in defining environments.} MAS environments are often abstracted or specified inconsistently~\cite{weyns2006environments}, which obscures whether observed cooperation arises from agent design, environmental affordances, or emergent properties.
    \item \textbf{Conflicting research agendas.} Hinder the evaluation and comparability of results~\cite{shoham2007if}, making it difficult to compare claims about cooperative behavior. 
    \item \textbf{Architectural-task inconsistency.} Some approaches define cooperation in terms of agent intent or architecture, while others define it purely in terms of task outcomes. For instance, in the Prisoner's Dilemma, even rule-based agents are said to ``cooperate'' if their behavior aligns with the game's cooperative payoff \cite{axelrod1981evolution}. This further blurs the boundary between coordination and cooperation in agent interpretation.
\end{enumerate}

The SMAC environment~\cite{samvelyan2019starcraft} provides a set of benchmark tasks for cooperative MARL, offering standardized challenge scenarios and best-practice guidelines. PettingZoo~\cite{terry2021pettingzoo} extends these efforts with a multi-agent environment library and a universal API to accelerate MARL research. More recently, Cross-environment Cooperation~\cite{jha2025cross} explores zero-shot multi-agent coordination. Earlier work by Fulda and Ventura~\cite{fulda2007predicting} highlighted the importance of predicting and preventing coordination problems in cooperative Q-learning systems. Our work complements such initiatives by addressing a preceding and often overlooked layer: the need for consistency at the meaning level. Instability in terms like \textit{cooperation} and \textit{coordination} across technical benchmarks risks reinforcing interpretive ambiguity rather than mitigating it.

Sun et al.~\cite{sun2025multi} survey coordination research and propose a unified three-component framework for coordination in sequential decision-making: system-level performance evaluation, social choice on who to coordinate with, and how to coordinate. While these models advance technical coordination, our work foregrounds an overlooked dimension — how semantic ambiguity, conceptual fluctuation, and moral overreadings surrounding \textit{cooperation} complicate the design and evaluation of MAS. 
 
Our goal is not to evaluate or craft definitions of \textit{coordination} and \textit{cooperation} but to highlight the urgency of the issues. Addressing this requires a shared conceptual vocabulary, a task that calls for a concerted multidisciplinary effort because \textit{cooperation}, as it manifests in MAS, crosses conceptual boundaries. It is not only a matter of policy or control, but of intention modeling, value projection, and interpretive inference, all of which are drawn from fields such as philosophy, cognitive science, and ethics. (This work acknowledges its own positioning within these evolving conceptual boundaries.)  



\section{The Misalignment Mosaic}\label{sec_mosaic}

In MAS design, misalignment can occur not only from agent behavior but also from gaps between layers of design, implementation, and interpretation. Assumptions made early in the design process (often informal, idealized, or morally charged) get embedded in the architecture, translated into behavior, and eventually perceived by stakeholders as intentional system properties. This layered fluctuation creates a critical question: where exactly did the misalignment originate? And who is accountable for its consequences?

Motivated by three interconnected drivers of meaning-level misalignment in MAS, the Misalignment Mosaic (depicted in Figure~\ref{fig:duck}, and detailed in the following subsections) provides a structured diagnostic framework to trace how misalignment emerges from language, framing, and epistemic assumptions, beyond technical mechanisms alone. Rather than resolve ambiguity through formal redefinition, the Misalignment Mosaic enables audits of meaning, illuminating where alignment falters at the conceptual level.

This framework builds on the concept of alignment magnification, which raises a critical question: how can AI systems be designed to pursue goals that remain consistent with their designers' intentions? Dung~\cite{dung2023current} warns that as AI systems become more capable, misalignment risks increase, making these systems not only more dangerous but potentially harder to align. Therefore, alignment should be considered foundational in developing MAS: ``No agent exists in a vacuum, and the evaluation of ethical behavior is a complex social and temporal phenomenon'' \cite{arnold2017value}.

In fact, the Morality-as-Cooperation Theory serves as a reminder that treating \textit{cooperation} as a black box introduces a critical source of misalignment in MAS. This conflation is not just a semantic detail, but a structural risk factor in how agent behavior is interpreted, evaluated, and morally framed. Addressing alignment must therefore proceed in parallel with refining our understanding of coordination and cooperation (and their variants). The Misalignment Mosaic integrates seemingly distinct issues to show their role in challenging alignment in MAS. When system designers fail to address ambiguities, agents may operate under misaligned assumptions, leading to goal failures, inefficiencies, or emergent behaviors that contradict human intent. Clarifying these terms is essential for designing MAS systems that operate reliably within both technical and societal constraints.

To support semantic audits and reflective analysis, we are developing a \textit{Mosaic Reflection Sheet}, an exploratory tool aligned with the Mosaic's components. The approach of Winikoff et al.~\cite{winikoff2025scoresheet}, who introduced a Scoresheet focused on explainability, inspired us. Their tool, applicable to MAS and broader AI technologies, incorporates stakeholder requirements and complements the IEEE P7001 Scoresheet~\cite{winfield2021ieee}, which focuses on transparency. The Mosaic Reflection Sheet may include one reflection sheet for each of the four Misalignment Mosaic components, all designed to support interpretive clarity and trace epistemic commitments across MAS development stages.

Finally, our efforts contribute to recent Cooperative AI priorities by enabling audits of semantic assumptions and conceptual fluctuations, which are increasingly necessary as MAS operate in norm-sensitive and socially embedded environments. We detail the Mosaic's four components next.

\subsection{Terminological Inconsistency}

The coordination/cooperation ambiguity leads to terminological inconsistencies, where terms are often defined interchangeably or imprecisely (see section~\ref{sec_background}). This fragmentation hinders theoretical and empirical progress, making it difficult to compare findings. Having \textit{cooperation} still largely treated as a black box is no surprise, given that design patterns for MAS are not very popular.

A review of the design patterns literature~\cite{juziuk2014design} reveals a notable absence of standardized templates, significantly impacting MAS practical application. This lack of uniformity results in inconsistent implementations, misinterpretations, and reduced transferability across systems. Furthermore, according to Juziuk and colleagues~\cite{juziuk2014design}, the relationships between patterns are poorly articulated, leading to a fragmented understanding of the pattern space: experts struggle to navigate and apply patterns effectively, limiting their practical utility. While MAS design patterns have been applied across various domains, existing classifications are largely confined to isolated catalogs, which prevents broader integration and adaptation~\cite{juziuk2014design}. A key example is the lack of standardized documentation and taxonomies for coordination and cooperation design patterns (along with variants). Without well-established standards, researchers apply similar terms to structurally distinct agent designs. While this issue also has interpretive consequences, we distinguish it from the Mosaic's Interpretive Ambiguity by anchoring it in terminological and design-level fragmentation. As MAS applications extend to large-scale deployment across critical industries such as healthcare, transportation systems, and food and agriculture \cite{keyIndustries}, the demand for systems that are not only efficient but also transparent and ethically grounded intensifies.

\subsection{Interpretive Ambiguity}

Illustrated by the Rabbit-Duck illusion, this component highlights ambiguities and interpretational limits in MAS research. Just as perception shifts in the illusion, MAS frameworks can be framed differently depending on theoretical or empirical perspectives, leading to inconsistent definitions of coordination and cooperation. In complex or uncertain environments, incomplete knowledge and partial observability further blur the distinction between coordinating tasks and cooperating for collective benefit, thereby amplifying the reach of the Misalignment Mosaic. While the Rabbit-Duck illusion broadly represents perceptual ambiguity, we use it here to illustrate interpretive shift between cooperation and coordination (and related terms), a shift that has conceptual and moral consequences in MAS design.

Serious issues can arise when algorithmic systems' goals are inferred solely from observations, potentially leading to misinterpretations of both the system's behavior and its intended objectives. Thus, ambiguity in defining cooperation frequently emerges in practice: consider agricultural robots tasked with harvesting crops. At first glance, their interactions may appear cooperative since they collectively contribute to the same task. However, closer analysis may reveal only coordination, where robots follow predefined movement patterns to avoid overlap rather than actively adapting to each other's presence or modifying strategies based on shared goals. Or, in another example, if various self-driving cars approach a four-way stop, should they coordinate by following predefined traffic rules or activate a cooperative mechanism that helps them dynamically yield? This ambiguity extends beyond system behavior to expert interpretation. Just as in the Rabbit-Duck illusion, it is possible that two researchers analyzing the same MAS testbed reach conflicting conclusions: one classifying it as a coordinative system, while the other sees cooperative behavior. This raises key epistemic and technical questions:

\begin{enumerate}   
 \item What defines a non-cooperative agent? Is it one that is inherently selfish, one that lacks an explicit cooperative mechanism, or something else?
\item Can cooperation emerge even when explicit cooperative mechanisms are absent? (What do we consider an explicit cooperative mechanism?)
\item How do experts (or users, if applicable) determine whether a MAS is coordinating or cooperating?
\end{enumerate}

As we try to address the questions above, we observe that multiple factors systematically influence whether MAS behavior is interpreted as coordinative or as indicative of cooperation. These factors span at least three levels of various considerations (note that \cite{huhns1999multiagent} provides insights into those):

\begin{itemize}
    \item \textbf{Level 1, Agent-level architecture and mechanisms}: perceptual mechanisms, goal structures, memory and learning processes, architectural constraints, rule-following implementations, input/output mappings, and action affordances.
    
    \item \textbf{Level 2, System-level emergent dynamics}: properties arising from interaction patterns, multi-agent dynamics, and communication potential. 
    
    \item \textbf{Level 3, Contextual and interpretive conditions}: task framing, deployment environment, and cross-disciplinary interpretive drift. 
\end{itemize}

Could a stakeholder interpret an agent's behavior as moral, cooperative, or aligned, even when it results from low-level coordination heuristics? What signaling mechanisms exist to distinguish surface behavior from intentional structure? Recognizing how these layers intersect is essential to addressing conceptual ambiguity. We argue that controlled experimentation, accompanied by clear documentation of these factors, is urgently needed to reduce interpretive drift and support more reliable evaluation of cooperation in MAS.

\subsubsection{If Coordination Exists, Is Cooperation Needed?}

Klassen et al.~\cite{klassen2023epistemic} demonstrate that partial knowledge can significantly impact agent beliefs and decision-making, sometimes leading to undesirable or even catastrophic outcomes. This raises a critical question for MAS: when agents face epistemic uncertainty, is coordination alone sufficient, or is cooperation necessary to mitigate these risks? In real-world deployments, agents rarely possess complete contextual knowledge due to limitations in data acquisition, processing, and environmental unpredictability.

While coordination alone may suffice in simpler, well-defined settings, real-world challenges require strategies that extend beyond simple coordination. Agents must navigate uncertainty, evolving conditions, and imperfect information, making cooperation a potentially necessary mechanism for adaptation and robustness. Given interpretive ambiguity, agents cannot rely solely on their knowledge and beliefs, as these limitations can adversely affect their decision-making capabilities (epistemic side effects~\cite{klassen2023epistemic}). This raises yet another question: if agents had perfect knowledge, would coordination alone be sufficient for optimal decision-making?

While theoretically possible, agents will operate under knowledge constraints in practice. Given this reality, could a cooperative model serve as a secondary mechanism, allowing agents to compensate for knowledge gaps and make more reliable decisions? This aligns with Curry \textit{et al.}~\cite{curry2019good}, who demonstrate that cooperative frameworks, grounded in shared goals and moral principles, offer effective approaches for addressing uncertainty and achieving collective success. Incorporating cooperative mechanisms into MAS may thus provide a more effective means of addressing epistemic limitations, ensuring more resilient decision-making in uncertain environments.

Focusing on distributed agent systems, Huhns and Stephens~\cite{huhns1999multiagent} treat agents collectively (as a society of agents) to address scalability challenges in environments with large agent populations. Their work presents a systematic approach to analyzing, describing, and designing multi-agent environments, taking into account both agent topology and content, which evolve dynamically. Notably, the authors~\cite{huhns1999multiagent} present a taxonomy of how agents coordinate behaviors and activities. In this framework, all elements (including cooperation and competition) emerge as forms of coordination. Coordination is seen as a system property, where agents operate within a shared environment, minimizing resource contention, avoiding livelock and deadlock, and maintaining safety conditions. In this taxonomy, cooperation is conceptualized as coordination among non-antagonistic agents, while negotiation represents coordination among competitive or self-interested agents. Then, successful cooperation typically would require agents to maintain models of one another and develop a model of future interactions, presupposing a degree of sociability.

A main takeaway from these considerations is that, to properly ask ``If coordination exists, is cooperation needed?'', the MAS community needs a clearer and consistent distinction between coordination and cooperation.

 \subsection{Concept to Code Decay}

As a MAS transitions from abstract concepts to operational structures, how can we ensure conceptual fidelity is maintained? Where are the translation gaps? What scaffolds (e.g., documentation, constraint encoding, behavior tracking) are in place to prevent conceptual fluctuation? Distortions may occur when theoretical models are translated into computational implementations, leading to unintended misalignments between the design intent and the real-world agents' behavior. As a system transitions from high-level conceptual design to implementation and deployment, critical nuances may be altered, omitted, or misinterpreted.

This can lead to a mismatch between initial goals (where \textit{cooperation} may have been clearly defined) and deployed systems (where cooperation degrades into partial or only coordination). As a result, Concept-to-Code Decay is a key factor within the broader Misalignment Mosaic.

 Knowledge exists at multiple levels of abstraction~\cite{abstractionTaxonomy}, and abstraction is vital in shaping both high-level conceptualization and specific design decisions in MAS. Hence, the level of abstraction decreases as the system transitions from conceptual design to concrete implementation, while its representativeness of the real-world application increases, often improving its practical feasibility but sometimes at the cost of oversimplifying key functional and ethical considerations. This transition reminds us of the Rabbit-Duck illusion: while a conceptual MAS model may capture intricate agent interactions and nuanced goals, its implementation may distort these complexities, either oversimplifying cooperative or coordinative mechanisms or introducing unintended emergent properties. This risks creating a gap between the system's observable behaviors and its intended behavioral goals, which constitutes a form of misalignment. As a result, concept-to-code decay can lead to misalignment, where the system's observable behaviors deviate from its intended design goals.

B{\u{a}}dic{\u{a}} and colleagues \cite{buadicua2018multi} inspire us to recognize that as abstraction decreases during the transition from conceptualization to deployment, key aspects of the original MAS motivation may be lost (concept-to-code decay). Addressing these abstraction gaps early in the design process is critical to ensuring that MAS implementations remain aligned with their intended objectives, minimizing both functional and ethical risks. Therefore, we must define terms and apply concepts carefully to align as much as possible within different stages, not only when informing a human about a system but also when designing, developing, testing, and assessing it. Misalignment can potentially lead to inefficiencies or even ethical concerns (especially when users are unaware of a system's underlying decision-making paradigms).

\begin{figure*}[h]
  \centering\includegraphics[width=0.5\linewidth]{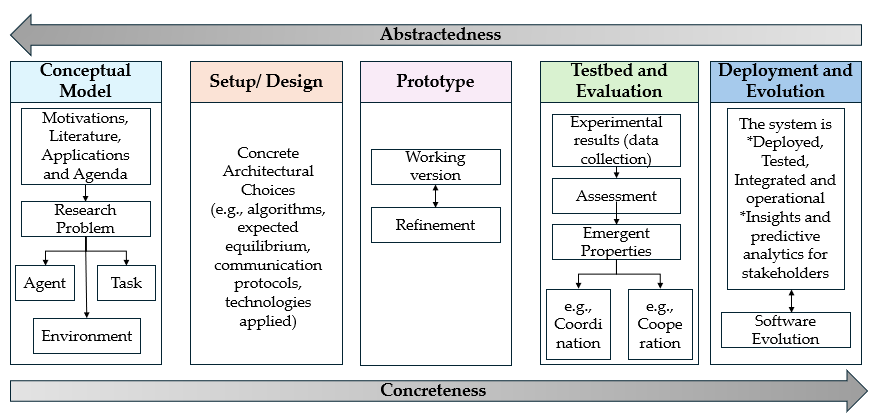}
  \caption{An illustration of a MAS transitioning from abstract concepts to concrete/operational.}
  \label{fig:AbLoss} \label{fig:abstraction}
\end{figure*}

In Figure~\ref{fig:abstraction}, we summarize the development and implementation of a MAS to illustrate the transition from abstract concepts to a concrete, operational entity. As the system progresses through stages -- Conceptual Model, Setup/Design, Prototype, Testbed/Evaluation, and Deployment/Evolution -- abstraction decreases while implementation details become more rigidly defined. However, as development advances, critical nuances (\textit{e.g.}, cooperative mechanisms, agent adaptability, and emergent behaviors) may be overlooked, leading to potential misalignment between the system's observable behavior and its original design goals. While this example treats coordination and cooperation as emergent properties discovered during evaluation, developers could have embedded them as explicit architectural choices early in the design phase. This flexibility underscores the importance of deliberate architectural decisions in shaping MAS behavior, ensuring that system alignment is proactively designed rather than emerging unpredictably. On a side note, Norval and colleagues \cite{norval2022disclosure} offer interesting insights into document engineering, while \cite{ElFassi02102024} into \textit{assumptions} using change propagation as context (they also reflect on \textit{assumption} definition); finally, UX/UI design also provides insights into reflecting on design assumptions.

\subsection{Morality-As-Cooperation Theory}

Achieving the best social outcome often requires agents to commit to cooperative strategies, \textit{i.e.}, agents must choose the actions that will only lead to the best outcome if most of the group commits and cooperates. However, despite its benefits, cooperation often results in a cost at the individual level, while defectors exploit shared resources \cite{wardil2014origin}. This dilemma between self-interest and collective maintenance arises in strategic settings, such as the Prisoner's Dilemma, and in real-world scenarios, including resistance to public health measures, where personal incentives undermine group stability.

From an evolutionary perspective, cooperation plays a central role in structuring group cohesion and maintaining social systems~\cite{tomasello2011human}. Tomasello and Vaish~\cite{tomasello2013origins} frame morality as an extension of cooperation, where pro-social motivations and cooperative behaviors scaffold the emergence of moral norms. Within this framing, cooperation requires alignment of individual and collective interests, often suppressing purely self-interested actions. Curry \textit{et al.}~\cite{curry2019good} extend this link through a cross-cultural lens, analyzing how 60 societies conceptualize cooperation. They investigate seven cooperative behaviors (coordination to mutual advantage among them) to test the Morality-as-Cooperation Theory. Their findings suggest that studied behaviors are strong candidates for universal moral rules, reinforcing the idea that morality is rooted in cooperative principles. Thus, the Morality-As-Cooperation Theory highlights the notion that \textit{cooperation} is deeply tied to moral rules, and understanding this connection is crucial for designing MAS that operate within ethical constraints.

Besides, it calls attention to \textit{cooperation} defined in terms of \textit{sacrificial cooperation} where self-interest or local utility is actively relinquished for the benefit of others, assuming no guaranteed reciprocity, and introducing loss, risk, or asymmetry into agent behavior, challenging utility-based models and exposing the limits of how cooperation is typically implemented in MAS. Thus, a question emerges: does cooperation involve individual costs or risks without assured gain (sacrificial), or is it based on mutual benefit and rational strategy (strategic)?

The distinction is crucial. In human societies, sacrificial cooperation (where individuals incur personal costs to uphold collective welfare, even absent guaranteed reciprocity) is central in moral systems~\cite{tomasello2013origins,wardil2014origin}. This raises fundamental questions about how MAS systems interpret, implement, and project cooperation, especially when deployed in environments where human observers may expect morally grounded, sacrificial behaviors — even when agents are only executing coordinative, utility-maximizing functions.

Still, in the MAS literature, cooperation is frequently framed as both a desirable behavior and a technical goal. Yet this apparent neutrality hides a deeper conflation: cooperation is often treated as a moral good, without acknowledging the ethical assumptions embedded in the term. In many MAS architectures, agents are labeled \textit{cooperative} when they optimize for shared outcomes, but these outcomes rarely undergo moral scrutiny. The system appears trustworthy not because it embodies moral reasoning, but because the label \textit{cooperative} carries inherent normative weight.

This blurring of cooperation and morality creates an interpretive shortcut: agents designed for coordination or strategic alignment may be perceived as morally aligned, regardless of their internal architectures or value systems. Such overreadings have real consequences, shaping trust, system evaluation, and normative governance in ways that extend beyond technical performance.

Recent work in the Foundations of Cooperative AI highlights these challenges \cite{conitzer2023foundations}. Conitzer and Oesterheld frame cooperation as a theoretical cornerstone for alignment, grounded in game-theoretic models. While we agree that conceptual clarity is essential, our work highlights a preceding issue: without diagnosing meaning-level misalignment, efforts to formalize cooperation risk overlooking critical zones that precede behavioral misalignment. This Mosaic's component (Morality-As-Cooperation Theory) urges the community \textbf{to recognize when technical alignment gets unintentionally morally charged.}


\section{Discussion}\label{sec_discussion}

MAS capabilities are widely acknowledged, as their architectures and applications offer unique advantages, which are particularly appealing for dynamic and distributed systems, enabling the efficient management of complex tasks and environments~\cite{kalachev2018intelligent}. For example, Liu et al.\cite{liu2019multi} emphasize the promise of MAS in optimizing the scheduling of aggregated resources and solving problems dynamically, demonstrating the transformative potential of these systems across domains.

Indeed, these technologies are already in active deployment, as noted by Müller \textit{et al.}~\cite{muller2014application}. Thus, Xue \textit{et al.}~\cite{xue2023rapid} describe a MAS-based control strategy for managing individual entities and bottom-layer control systems, highlighting MAS's potential to revolutionize industries through improved coordination, scalability, and adaptability. However, as deployment scenarios become more complex, the distinction between coordination and cooperation becomes operationally indistinct under real-world conditions. Without standardized frameworks, MAS models risk misclassifying agent behaviors, leading to inefficiencies and unintended emergent properties.

The Rabbit-Duck illusion exemplifies how the same visual input can yield different interpretations. Similarly, agent behaviors in MAS can be perceived as either cooperative or only coordinative, depending on theoretical and methodological perspectives. Therefore, the ambiguity between coordination and cooperation complicates MAS evaluation, necessitating clearer guidelines and standardized definitions. Recent work by Kierans et al.~\cite{kierans2024} highlights the need to frame alignment challenges from a sociotechnical perspective, incorporating multiple human and AI agents. While our focus remains on conceptual instability within MAS, their call reinforces that misalignment is layered, emerging not only from technical failures but from interpretive ambiguity embedded in system design and evaluation.

We introduced the Misalignment Mosaic, a diagnostic framework that articulates how overlooked semantic and interpretative distinctions (especially between cooperation and coordination) contribute to misalignment in MAS. We list below some implications of the Misalignment Mosaic:
\begin{enumerate}
    \item Recognizing misalignment as a multi-layered issue makes it clear that no single intervention can fully resolve systemic failures. A mosaic perspective compels experts to assess the entire ecosystem of MAS (spanning from meaning and conceptual definitions, benchmarking practices, and real-world interpretability) rather than treating misalignment as a distant problem.    
    
 \item Concept-to-Code Decay and Early-Stage Alignment Checks: conceptual models may lose key properties during implementation and deployment, resulting in reduced representativeness. This distorts the intended functionality and raises ethical concerns, especially when algorithmic systems interact with humans. Therefore, by viewing misalignment as a network of interlinked issues, designers are encouraged to integrate conceptual clarity into the early stages of development. This means explicitly defining coordination and cooperation before they manifest in code, ensuring that user-facing interfaces reflect intended system behaviors rather than assumed ones.
       
\item Refinement of terminology and benchmarks: without structured documentation and classification, the MAS community faces significant barriers to aligning with both technical and moral goals. The Misalignment Mosaic underscores the necessity for standardized definitions and clear evaluation frameworks that differentiate coordination from cooperation. Otherwise, both experimental results and real-world deployments risk misinterpreting system alignment and, therefore, leading to inefficiencies, failure cases, or unverified claims about agent behavior.

\item Urgency: identified challenges are now scaling to agentic workflows, where coordination and cooperation ambiguities become even more dangerous. Large-scale industrial deployments demand trust, reliability, and interpretability.
\end{enumerate}

\textbf{Call for action.} We urge the community to partner with other disciplines and undertake a systematic investigation of terminology across the literature and: 1) Develop a standardized MAS taxonomy for coordination, cooperation, and variants. 2) Create dynamic platforms and benchmarks for evaluating those under varied conditions. These platforms would provide the necessary scaffolding to assess and improve coordination and cooperation mechanisms. Finally, 3) Integrate sociotechnical perspectives: expanding alignment research to multidisciplinary and multicultural considerations. This reinforces the idea that cooperation and coordination are not just technical constructs but integral to both moral and functional systems~\cite{curry2019good}, and vital to other disciplines as well.

\textbf{Limitations and Future Directions.} This work adopts a conceptual and diagnostic approach to misalignment in MAS, focusing on the meaning-level and interpretive conditions that precede technical failures. While we present the Misalignment Mosaic, we must acknowledge its limitations: our analysis centers on the ambiguity between cooperation and coordination. While we argue that this ambiguity is foundational to many misalignment risks, other overloaded terms in MAS (such as autonomy, alignment, or trust) require parallel diagnostic attention. Future efforts should explore how semantic instability propagates across AI domains, particularly as MAS-like architectures underpin increasingly complex agentic systems. 

Despite these limitations, we believe this work offers a necessary foundation for auditing meaning as a structural dimension of alignment, a dimension often overlooked in technical models of MAS design and governance.

\section{Conclusion}\label{sec_conclusion}

We introduced the Misalignment Mosaic as a framework for diagnosing and tracing meaning-level misalignment in MAS. This work does not propose an empirical resolution to this ambiguity, but rather calls for a structural shift in how the MAS community frames, audits, and addresses meaning-level alignment. Conceptual clarity must precede empirical validation; as long as foundational MAS concepts remain unstable, no amount of technical refinement will ensure alignment at scale.

While our discussion draws from existing empirical studies in MAS and MARL, we emphasize that the next step is not immediate experimentation, but a coordinated, multidisciplinary effort to establish clearer terminology, benchmarks, and conceptual scaffolding. Clarifying meaning is a prerequisite for building aligned, trustworthy algorithmic systems.

We invite the community to reflect: is there a connection between misalignment and moral overreadings with treating cooperation as a black box? If we seek to build AI systems that cooperate \textit{purposefully} with humans and across cultural boundaries, it is vital to establish shared conceptual scaffolds that define cooperation across disciplinary, institutional, and deployment contexts.

\section*{Acknowledgments}
This work would not have been possible without the support from Grinnell College's Harris Faculty Fellowship and the Mentored Advanced Project (MAP) program. We would like to thank our lab members Alyssa Trapp and Jio Hong for asking the question: "Is cooperation needed if coordination exists?"

The authors made use of generative AI tools (ChatGPT, OpenAI) for language refinement and structural suggestions. All intellectual framing, conceptual decisions, and conclusions are the authors' own.

\bibliographystyle{unsrt}  
\bibliography{references}  

@article{ElFassi02102024,
author = {Soufiane El Fassi and Xin Chen and Atif Riaz and Marin D. Guenov and Albert S.J. van Heerden and Sergio Jimeno Altelarrea},
title = {Managing assumption-driven design change via margin allocation and trade-offs},
journal = {Journal of Engineering Design},
volume = {35},
number = {10},
pages = {1258--1291},
year = {2024},
publisher = {Taylor \& Francis},
doi = {10.1080/09544828.2023.2259741},
URL = {https://doi.org/10.1080/09544828.2023.2259741},
eprint = {https://doi.org/10.1080/09544828.2023.2259741}
}

@inproceedings{norval2022disclosure,
  title={Disclosure by Design: Designing information disclosures to support meaningful transparency and accountability},
  author={Norval, Chris and Cornelius, Kristin and Cobbe, Jennifer and Singh, Jatinder},
  booktitle={Proceedings of the 2022 ACM Conference on Fairness, Accountability, and Transparency},
  pages={679--690},
  year={2022}
}

@article{tomasello2011human,
  title={Human culture in evolutionary perspective},
  author={Tomasello, Michael},
  journal={Advances in culture and psychology},
  volume={1},
  pages={5--51},
  year={2011},
  publisher={Oxford University Press New York, NY}
}

@article{tomasello2013origins,
  title={Origins of human cooperation and morality},
  author={Tomasello, Michael and Vaish, Amrisha},
  journal={Annual review of psychology},
  volume={64},
  pages={231--255},
  year={2013},
  publisher={Annual Reviews}
}

@article{winfield2021ieee,
  title={IEEE P7001: A proposed standard on transparency},
  author={Winfield, Alan FT and Booth, Serena and Dennis, Louise A and Egawa, Takashi and Hastie, Helen and Jacobs, Naomi and Muttram, Roderick I and Olszewska, Joanna I and Rajabiyazdi, Fahimeh and Theodorou, Andreas and others},
  journal={Frontiers in Robotics and AI},
  volume={8},
  pages={665729},
  year={2021},
  publisher={Frontiers Media SA}
}

@article{shaikh2020braitenberg,
  title={Braitenberg vehicles as computational tools for research in neuroscience},
  author={Shaikh, Danish and Ra{\~n}{\'o}, Ignacio},
  journal={Frontiers in bioengineering and biotechnology},
  volume={8},
  pages={565963},
  year={2020},
  publisher={Frontiers Media SA}
}

@incollection{prescott2019synthetic,
  title={The synthetic psychology of the self},
  author={Prescott, T. J. and Camilleri, D.},
  booktitle={Cognitive architectures},
  pages={85--104},
  year={2019},
  publisher={Springer}
}

@inproceedings{fulda2007predicting,
  title={Predicting and Preventing Coordination Problems in Cooperative Q-learning Systems.},
  author={Fulda, Nancy and Ventura, Dan},
  booktitle={IJCAI},
  volume={2007},
  pages={780--785},
  year={2007}
}

@inproceedings{duque2024advantage,
title={Advantage Alignment Algorithms},
author={Juan Agustin Duque and Milad Aghajohari and Tim Cooijmans and Tianyu Zhang and Aaron Courville},
booktitle={ICML 2024 Workshop: Aligning Reinforcement Learning Experimentalists and Theorists},
year={2024},
url={https://openreview.net/forum?id=XK5981wGkW}
}

@article{jha2025cross,
  title={Cross-environment Cooperation Enables Zero-shot Multi-agent Coordination},
  author={Jha, Kunal and Carvalho, Wilka and Liang, Yancheng and Du, Simon S and Kleiman-Weiner, Max and Jaques, Natasha},
  journal={arXiv preprint arXiv:2504.12714},
  year={2025}
}

@inproceedings{terry2021pettingzoo,
  author = {Terry, J and Black, Benjamin and Grammel, Nathaniel and Jayakumar, Mario  and Hari, Ananth  and Sullivan, Ryan and Santos, Luis S and Dieffendahl, Clemens and Horsch, Caroline and Perez-Vicente, Rodrigo and Williams, Niall  and Lokesh, Yashas  and Ravi, Praveen },
 booktitle = {Advances in Neural Information Processing Systems},
 editor = {M. Ranzato and A. Beygelzimer and Y. Dauphin and P.S. Liang and J. Wortman Vaughan},
 pages = {15032--15043},
 publisher = {Curran Associates, Inc.},
 title = {PettingZoo: Gym for Multi-Agent Reinforcement Learning},
 url = {https://proceedings.neurips.cc/paper_files/paper/2021/file/7ed2d3454c5eea71148b11d0c25104ff-Paper.pdf},
 volume = {34},
 year = {2021}
}

@article{samvelyan2019starcraft,
  title={The starcraft multi-agent challenge},
  author={Samvelyan, Mikayel and Rashid, Tabish and De Witt, Christian Schroeder and Farquhar, Gregory and Nardelli, Nantas and Rudner, Tim GJ and Hung, Chia-Man and Torr, Philip HS and Foerster, Jakob and Whiteson, Shimon},
  journal={arXiv preprint arXiv:1902.04043},
  year={2019}
}

@inproceedings{winikoff2025scoresheet,
  title={A Scoresheet for Explainable AI},
  author={Winikoff, Michael and Thangarajah, John and Rodriguez, Sebastian},
  booktitle={Proc. of the 24th International Conference on Autonomous Agents and Multiagent Systems},
  pages={2171--2180},
  year={2025}
}

@misc{coordination_def,
	title = {Coordination},
author = {Merriam-Webster},
	url = {https://www.merriam-webster.com/dictionary/coordination#:~:text=1,requires%20excellent%20hand%2Deye%20coordination.},
	urldate = {2024-10-20},
        year={2024},
	publisher={Merriam-Webster.com}
}

@misc{concept_def,
	title = {Concept},
author = {Merriam-Webster},
	url = {https://www.merriam-webster.com/dictionary/concept},
	urldate = {2025-07-24},
        year={2025},
	publisher={Merriam-Webster.com}
}

@misc{term_def,
	title = {Term},
author = {Merriam-Webster},
	url = {https://www.merriam-webster.com/dictionary/term},
	urldate = {2025-07-24},
        year={2025},
	publisher={Merriam-Webster.com}
}

@misc{coop_def,
	title = {Cooperation},
author = {Merriam-Webster},
	url = {https://www.merriam-webster.com/dictionary/cooperation},
	urldate = {2024-10-20},
        year={2024},
	publisher={Merriam-Webster.com}
}

@article{matignon2012independent,
  title={Independent reinforcement learners in cooperative Markov games: a survey regarding coordination problems},
  author={Matignon, Laetitia and Laurent, Guillaume J and Le Fort-Piat, Nadine},
  journal={The Knowledge Engineering Review},
  volume={27},
  number={1},
  pages={1--31},
  year={2012},
  publisher={Cambridge University Press}
}

@inproceedings{boutilier1999sequential,
  title={Sequential optimality and coordination in multiagent systems},
  author={Boutilier, Craig},
  booktitle={IJCAI'99: Proceedings of the 16th international joint conference on Artificial intelligence},
  volume={1},
  pages={478--485},
  year={1999}
}

@inproceedings{erculiani2023egocentric,
  title={Egocentric Hierarchical Visual Semantics},
  author={Erculiani, Luca and Bontempelli, Andrea and Passerini, Andrea and Giunchiglia, Fausto and others},
  booktitle={Proceedings of the Second International Conference on Hybrid Human-Machine Intelligence},
  year={2023},
  organization={IOS Press}
}

@book{wittgenstein,
  title={Philosophical Investigations},
  author={Ludwig Wittgenstein},
  year={1953},
  publisher={}
}

@book{braitenberg1986vehicles,
  title={Vehicles: Experiments in synthetic psychology},
  author={Braitenberg, Valentino},
  year={1986},
  publisher={MIT press}
}

@article{gibson1966senses,
  title={The senses considered as perceptual systems.},
  author={Gibson, James Jerome},
  year={1966},
  publisher={Houghton Mifflin}
}

@article{wardil2014origin,
  title={Origin and structure of dynamic cooperative networks},
  author={Wardil, Lucas and Hauert, Christoph},
  journal={Scientific reports},
  volume={4},
  number={1},
  pages={5725},
  year={2014},
  publisher={Nature Publishing Group UK London}
}

@book{gibson2014ecological,
  title={The ecological approach to visual perception: classic edition},
  author={Gibson, James Jerome},
  year={[1979], 2014},
  publisher={Psychology press}
}

@inproceedings{Vehicles2023,
author = {Swaim, Elliot and Eliott, Fernanda},
title = {Complex Behavior Vs. Design – Interpreting AI: Reminders from Synthetic Psychology},
year = {2023},
publisher = {Iaria},
booktitle = {The Ninth International Conference on Human and Social Analytics HUSO, Barcelona, Spain.}
}

@inproceedings{weyns2006environments,
  title={Environments for situated multi-agent systems: Beyond infrastructure},
  author={Weyns, Danny and Vizzari, Giuseppe and Holvoet, Tom},
  booktitle={Environments for Multi-Agent Systems II: Second International Workshop, E4MAS 2005, Utrecht, The Netherlands, July 25, 2005, Selected Revised and Invited Papers 2},
  pages={1--17},
  year={2006},
  organization={Springer}
}

@book{barad2007meeting,
  title={Meeting the universe halfway: Quantum physics and the entanglement of matter and meaning},
  author={Barad, Karen},
  year={2007},
  publisher={duke university Press}
}

@article{mesly2022terminology,
  title={Terminology Matters: A Review on the Concept of Economic Predation},
  author={Mesly, Olivier and Petrescu, Maria and Mesly, Alexandra},
  journal={Journal of Economic Issues},
  volume={56},
  number={4},
  pages={959--987},
  year={2022},
  publisher={Taylor \& Francis}
}

@article{eitzel2017citizen,
  title={Citizen science terminology matters: Exploring key terms},
  author={Eitzel, Melissa and Cappadonna, Jessica and Santos-Lang, Chris and Duerr, Ruth and West, Sarah Elizabeth and Virapongse, Arika and Kyba, Christopher and Bowser, Anne and Cooper, Caren and Sforzi, Andrea and others},
  journal={Citizen science: Theory and practice},
  pages={1--20},
  year={2017},
  publisher={York}
}

@inproceedings{langer2022look,
  title={``Look! It's a computer program! It's an algorithm! It's AI!'': Does terminology affect human perceptions and evaluations of algorithmic decision-making systems?},
  author={Langer, Markus and Hunsicker, Tim and Feldkamp, Tina and K{\"o}nig, Cornelius J and Grgi{\'c}-Hla{\v{c}}a, Nina},
  booktitle={Proceedings of the 2022 CHI Conference on Human Factors in Computing Systems},
  pages={1--28},
  year={2022}
}

@article{shoham2007if,
  title={If multi-agent learning is the answer, what is the question?},
  author={Shoham, Yoav and Powers, Rob and Grenager, Trond},
  journal={Artificial intelligence},
  volume={171},
  number={7},
  pages={365--377},
  year={2007},
  publisher={Elsevier}
}

@article{huhns1999multiagent,
  title={Multiagent systems and societies of agents},
  author={Huhns, Michael N and Stephens, Larry M},
  journal={Multiagent systems: a modern approach to distributed artificial intelligence},
  volume={1},
  pages={79--114},
  year={1999},
  publisher={MIT press Cambridge, MA}
}

@article{tuyls2012multiagent,
  title={Multiagent learning: Basics, challenges, and prospects},
  author={Tuyls, Karl and Weiss, Gerhard},
  journal={Ai Magazine},
  volume={33},
  number={3},
  pages={41--41},
  year={2012}
}

@inproceedings{conitzer2023foundations,
  title={Foundations of cooperative AI},
  author={Conitzer, Vincent and Oesterheld, Caspar},
  booktitle={Proceedings of the AAAI Conference on Artificial Intelligence},
  volume={37},
  number={13},
  pages={15359--15367},
  year={2023}
}

@inproceedings{zhang2016formal,
  title={A formal analysis of required cooperation in multi-agent planning},
  author={Zhang, Yu and Sreedharan, Sarath and Kambhampati, Subbarao},
  booktitle={Proceedings of the International Conference on Automated Planning and Scheduling},
  volume={26},
  pages={335--343},
  year={2016}
}

@article{axelrod1981evolution,
  title={The evolution of cooperation},
  author={Axelrod, Robert and Hamilton, William D},
  journal={science},
  volume={211},
  number={4489},
  pages={1390--1396},
  year={1981},
  publisher={American Association for the Advancement of Science}
}

@inproceedings{tan1993multi,
  title={Multi-agent reinforcement learning: Independent vs. cooperative agents},
  author={Tan, Ming},
  booktitle={Proceedings of the tenth international conference on machine learning},
  pages={330--337},
  year={1993}
}

@inproceedings{consoli2007link,
  title={The link between agent coordination and cooperation},
  author={Consoli, Angela and Tweedale, Jeffrey and Jain, Lakhmi},
  booktitle={Intelligent Information Processing III: IFIP TC12 International Conference on Intelligent Information Processing (IIP 2006), September 20--23, Adelaide, Australia 3},
  pages={11--19},
  year={2007},
  organization={Springer}
}

@inproceedings{malone1990coordination,
  title={What is coordination theory and how can it help design cooperative work systems?},
  author={Malone, Thomas W and Crowston, Kevin},
  booktitle={Proceedings of the 1990 ACM conference on Computer-supported cooperative work},
  pages={357--370},
  year={1990}
}

@article{jensen2015testing,
  title={Testing theories about ethnic markers: Ingroup accent facilitates coordination, not cooperation},
  author={Jensen, Niels Holm and Petersen, Michael Bang and H{\o}gh-Olesen, Henrik and Ejstrup, Michael},
  journal={Human Nature},
  volume={26},
  pages={210--234},
  year={2015},
  publisher={Springer}
}

@book{lynch1996distributed,
  title={Distributed Algorithms},
  author={Lynch, Nancy A.},
  year={1996},
  publisher={Morgan Kaufmann Publishers}
}

@book{chakravarty2014course,
  title={A course on cooperative game theory},
  author={Chakravarty, Satya R and Mitra, Manipushpak and Sarkar, Palash},
  year={2014},
  publisher={Cambridge University Press}
}

@article{lanctot2017unified,
  title={A unified game-theoretic approach to multiagent reinforcement learning},
  author={Lanctot, Marc and Zambaldi, Vinicius and Gruslys, Audrunas and Lazaridou, Angeliki and Tuyls, Karl and P{\'e}rolat, Julien and Silver, David and Graepel, Thore},
  journal={Advances in neural information processing systems},
  volume={30},
  year={2017}
}

@article{du2023review,
  title={A Review of Cooperation in Multi-agent Learning},
  author={Du, Yali and Leibo, Joel Z and Islam, Usman and Willis, Richard and Sunehag, Peter},
  journal={arXiv e-prints},
  pages={arXiv--2312},
  year={2023}
}

@article{abstractionTaxonomy,
 ISSN = {17456916, 17456924},
 URL = {http://www.jstor.org/stable/26358684}, 
 author = {Stephen K. Reed},
 journal = {Perspectives on Psychological Science},
 number = {6},
 pages = {817--837},
 publisher = {[Association for Psychological Science, Sage Publications, Inc.]},
 title = {A Taxonomic Analysis of Abstraction},
 urldate = {2024-06-21},
 volume = {11},
 year = {2016}
}

@misc{keyIndustries,
author = {{America's Cyber Defense Agency}},
title = {Critical Infrastructure Sectors},
howpublished = {\url{https://www.cisa.gov/topics/critical-infrastructure-security-and-resilience/critical-infrastructure-sectors}},
note = {Accessed: 02/25/2025}
}

@article{hanks1993benchmarks,
  title={Benchmarks, test beds, controlled experimentation, and the design of agent architectures},
  author={Hanks, Steve and Pollack, Martha E and Cohen, Paul R},
  journal={AI magazine},
  volume={14},
  number={4},
  pages={17--17},
  year={1993}
}

@article{sun2025multi,
  title={Multi-agent coordination across diverse applications: A survey},
  author={Sun, Lijun and Yang, Yijun and Duan, Qiqi and Shi, Yuhui and Lyu, Chao and Chang, Yu-Cheng and Lin, Chin-Teng and Shen, Yang},
  journal={arXiv preprint arXiv:2502.14743},
  year={2025}
}

@article{curry2019good,
  title={Is it good to cooperate? Testing the theory of morality-as-cooperation in 60 societies},
  author={Curry, Oliver Scott and Mullins, Daniel Austin and Whitehouse, Harvey},
  journal={Current anthropology},
  volume={60},
  number={1},
  pages={47--69},
  year={2019},
  publisher={The University of Chicago Press Chicago, IL}
}

@article{buadicua2018multi,
  title={Multi-agent modelling and simulation of graph-based predator--prey dynamic systems: A BDI approach},
  author={B{\u{a}}dic{\u{a}}, Amelia and B{\u{a}}dic{\u{a}}, Costin and Ivanovi{\'c}, Mirjana and D{\u{a}}nciulescu, Daniela},
  journal={Expert Systems},
  volume={35},
  number={5},
  pages={e12263},
  year={2018},
  publisher={Wiley Online Library}
}

@inproceedings{klassen2023epistemic,
  title={Epistemic side effects: An AI safety problem},
  author={Klassen, Toryn Q and Alamdari, Parand Alizadeh and McIlraith, Sheila A},
  booktitle={Proceedings of the 2023 International Conference on Autonomous Agents and Multiagent Systems},
  pages={1797--1801},
  year={2023}
}

@article{juziuk2014design,
  title={Design patterns for multi-agent systems: A systematic literature review},
  author={Juziuk, Joanna and Weyns, Danny and Holvoet, Tom},
  journal={Agent-oriented software engineering: reflections on architectures, methodologies, languages, and frameworks},
  pages={79--99},
  year={2014},
  publisher={Springer}
}

@inproceedings{kalachev2018intelligent,
  title={Intelligent mechatronic system with decentralised control and multi-agent planning},
  author={Kalachev, Andrei and Zhabelova, Gulnara and Vyatkin, Valeriy and Jarvis, Dennis and Pang, Cheng},
  booktitle={IECON 2018-44th Annual Conference of the IEEE Industrial Electronics Society},
  pages={3126--3133},
  year={2018},
  organization={IEEE}
}

@article{liu2019multi,
  title={A multi-agent architecture for scheduling in platform-based smart manufacturing systems},
  author={Liu, Yong-kui and Zhang, Xue-song and Zhang, Lin and Tao, Fei and Wang, Li-hui},
  journal={Frontiers of Information Technology \& Electronic Engineering},
  volume={20},
  number={11},
  pages={1465--1492},
  year={2019},
  publisher={Springer}
}

@article{xue2023rapid,
  title={Rapid deployment scheme for MAS-based applications within industry 4.0 framework},
  author={Xue, Liwei and Liu, Guo-Ping and Hu, Wenshan},
  journal={IFAC-PapersOnLine},
  volume={56},
  number={2},
  pages={3850--3855},
  year={2023},
  publisher={Elsevier}
}

@article{muller2014application,
  title={Application impact of multi-agent systems and technologies: A survey},
  author={M{\"u}ller, J{\"o}rg P and Fischer, Klaus},
  journal={Agent-Oriented Software Engineering: Reflections on Architectures, Methodologies, Languages, and Frameworks},
  pages={27--53},
  year={2014},
  publisher={Springer}
}

@misc{kierans2024,
      title={Quantifying Misalignment Between Agents: Towards a Sociotechnical Understanding of Alignment}, 
      author={Aidan Kierans and Avijit Ghosh and Hananel Hazan and Shiri Dori-Hacohen},
      year={2024},
      eprint={2406.04231},
      archivePrefix={arXiv},
      primaryClass={cs.MA},
      url={https://arxiv.org/abs/2406.04231}, 
}

@inproceedings{arnold2017value,
  title={Value Alignment or Misalignment--What Will Keep Systems Accountable?},
  author={Arnold, Thomas and Kasenberg, Daniel and Scheutz, Matthias},
  booktitle={Workshops at the thirty-first AAAI conference on artificial intelligence},
  year={2017}
}

@article{dung2023current,
  title={Current cases of AI misalignment and their implications for future risks},
  author={Dung, Leonard},
  journal={Synthese},
  volume={202},
  number={5},
  pages={138},
  year={2023},
  publisher={Springer}
}

\end{document}